\documentclass{llncs}

\usepackage{graphicx}
\usepackage{amsfonts}
\usepackage{amsmath}  
\usepackage{latexsym} 
\usepackage{amssymb}
 
\newcommand{\ora}[1]{\overrightarrow{#1}}

\newcommand{\ket}[1]{| #1 \rangle}
\newcommand{\bra}[1]{\langle #1 |}

\newtheorem{Def}{Definition}

\newtheorem{Wdef}{Working definition}
\newtheorem{Th}{Theorem}

\newtheorem{Lem}{Lemma}
\newtheorem{Proposition}{Proposition}
\newtheorem{Rk}{Remark}
\newtheorem{Open}{Open problem}

\begin{document}

\title{
Algebraic Characterizations of Unitary Linear Quantum Cellular Automata
}

\author{Pablo Arrighi}

\institute{IMAG Laboratories \& University of Grenoble,\\ 
46 Avenue F\'elix Viallet, 38031 Grenoble Cedex, France. \\
\email{pablo.arrighi@imag.fr}}

\maketitle

\begin{abstract}
We provide algebraic criteria for the unitarity of
linear quantum cellular automata, i.e. one dimensional 
quantum cellular automata. We derive these both
by direct combinatorial arguments, and by adding 
constraints into the model which do not change the quantum 
cellular automata's computational power. The configurations 
we consider have finite but unbounded size.
\end{abstract}


\section{Motivations}\label{sectionmotivations}
One could say that the central question in theoretical 
computer science is `What are the resources necessary for 
computation, or information processing?'. Ultimately this 
question is dictated by the physical laws which surround 
us. Quantum computation science has risen from this basic 
idea. It considers computers as physical, and hence 
possibly quantum systems. At the theoretical level it was 
demonstrated for instance that polynomial-time integer 
factorization is possible with such systems, as well as the 
search for an element in a unordered list of size $n$ in 
time $\Theta(\sqrt{n})$.\\ Cellular automata (CA) are 
arrays of cells, each of which may take one in a finite 
number of possible state. These evolve in discrete time 
steps according to a global evolution $\Delta$ -- which 
itself arises from the application of a local transition 
function $\delta$, synchronously and homogeneously across 
space. A popular example is Conway's `Game of Life', a two 
dimensional CA which has been proven to be universal for 
computation.\\ It is clear that CA are themselves 
physics-like models of computations, as they describe a
world of small systems interacting locally, according to
translation-invariant laws. Therefore it seems 
natural to study their quantum extensions. Moreover no 
classical control is required in such models, since 
computation arises as an emergent behaviour of the quantum 
cells' interaction. This is a key advantage to have as it
reduces the need for environment interaction, and hence may
reduce decoherence, the principal obstacle to realizing a
quantum computer. For these two reasons Feynman 
\cite{Feynman}, in his seminal paper about quantum computation, 
has argued that the study of quantum CA may 
prove an important path to a realistic physical 
implementation of quantum computers. Another series of
legitimate aims is to endow quantum computation with 
spatio-temporal notions, or even to provide a bridge through which computer 
science notions, such as universality, may contribute to 
modern theoretical physics. To put it differently such works
are a contribution to the understanding of dynamics in
discrete, quantum spacetime -- but from an idealized,
computer-science viewpoint.\\
Similarly to their classical counterparts linear 
quantum cellular automata (LQCA) consist of a row of 
identical, finite dimensional, quantum systems. These 
evolve in discrete time steps according to a global 
evolution $\Delta$ --  which itself arises from the 
application of a local transition function $\delta$, 
homogeneously and synchronously across space. But in order 
to grant LQCA the status of physically acceptable model of 
computation, one must ensure that the global evolution 
$\Delta$ is physically acceptable in a quantum theoretical 
setting, i.e. one must ensure that $\Delta$ is unitary. 
Unfortunately this global property is rather non-trivially 
related to the description of the local transition function 
$\delta$ -- witness of this the abundant literature on 
reversible cellular automata (RCA), tackling the classical 
counterpart of this issue. It is actually a very surprising
fact that so much has been done to study RCA 
-- when reversibility is not so much of a crucial feature 
to have in classical computation. A frequently encountered 
argument states that all consumption-less, zero-heat 
micro-mechanical device need be reversible. But tracing back 
the origins of this argument, we find quantum physical 
considerations once more \cite{Landauer}.\\ 
One way to approach this issue is to 
find a decision procedure, which given $\delta$ tells 
whether $\Delta$ is unitary. This test should be performed 
efficiently, so as not to carry any of the complexity of 
the computation, and be applied to any candidate $\delta$ 
as a mean to exclude the non-physical ones. Such a strong 
contribution was indeed achieved by D\"urr et al. It does 
not put an ending point to the problem however, because the 
number of local transition functions which do indeed induce 
a unitary global evolution is likely to be rather scarce, 
as was the case for RCA \cite{Amoroso}. Moreover this 
relatively complicated decision procedure takes an elegant 
detour via finite automata, but one which does not guide us 
to understand which $\delta$'s will eventually yield a 
physical $\Delta$.\\ Physicists are used to checking 
whether an evolution is physically acceptable, but they 
like to do so algebraically (e.g. 
$U^{\dagger}U=UU^{\dagger}=\mathbb{I}$ for unitarity, when 
$U$ is a finite matrix). Much in the same way computer 
scientists are used to checking whether a program is valid, 
but like these criteria to be syntactic. When this is not 
the case, we tend to consider that the definition chosen to 
formalize the model of computation is in fact too lose. 
Indeed once universality has been reached, adding more 
expressiveness does not mean adding more computational power, 
but only more ways of expressing the same computation. An 
undesirable excess arises when the syntax proposed by the 
definition allows the description of non-valid 
(unrealistic, non-physical) programs, thereby requiring 
that the user performs non-trivial (non-syntactic) decision 
procedures to exclude those instances. This is the current 
state of affairs with LQCA.\\ Therefore a different, 
complementary approach is to tighten the definition of linear quantum cellular automata, i.e. to seek for a 
more restrictive definition whose unitarity may be checked 
algebraically/syntactically, and yet capable of expressing 
the exact same set of global evolutions as our original 
definition. In this paper we provide algebraic 
characterizations of unitary linear quantum 
cellular automata. We derive these both
by direct combinatorial arguments, and by adding 
constraints into the model which do not change the quantum 
cellular automata's computational power.\\ 
The breakdown of the core of this 
paper will be given after LQCA are presented formally,  in 
the following section. Our main theorem is stated in the 
conclusion section, and discussed in comparison to some 
related approaches.\\
\noindent\emph{Notations.} Throughout the paper we will 
denote by $\mathcal{H}_S$ the hilbert space whose canonical 
orthonormal base vectors are identified with the elements 
of the countable set $S$. E.g. $\mathcal{H}_{\{aa, ab, ba, 
bb\}}$ is the four dimensional space with canonical 
orthonormal base $\{\ket{aa}, \ket{ab}, \ket{ba}, 
\ket{bb}\}$. This means that any vector 
$\alpha\ket{aa}+\beta\ket{ab}+\gamma\ket{ba}+\delta\ket{bb}$ with 
$\alpha, \beta, \gamma, \delta\in \mathbb{C}$ belongs to 
$\mathcal{H}_{\{aa, ab, ba, bb\}}$. Such a vector must be 
thought of as a superposition of the words $aa, ab, ba, 
bb$. Moreover the symbol $\mathbf{0}$ is to denote the null 
vector, not to be confused with $O$ the matrix containing 
only ones.

\section{The model}\label{sectionmodel}

\noindent We start with the definition proposed by 
\cite{Watrous} and \cite{Durr}. This definition 
will evolve throughout the paper. 
\begin{Wdef}[LQCA]~\label{lqca}\\ A linear quantum cellular 
automaton (LQCA) is a 4-tuple $\mathcal{A}=(\Sigma, q, N, 
\delta)$, where (with $q\Sigma=\{q\}\cup\Sigma$):\vspace{1mm}\\
- $\Sigma$ is a finite set of symbols (i.e. ``the 
alphabet", giving the possible basic states each cell may 
take);\\
- $q$ is a symbol such that $q\notin\Sigma$ (i.e. ``the 
quiescent symbol", which may be thought as a special state 
for empty cells);\\
- $N$ is a set of n successive signed integers (i.e. 
``the neighbourhood", telling which cell is next to whom);\\
- $\delta: \mathcal{H}_{(q\Sigma)^{n}}\rightarrow 
\mathcal{H}_{q\Sigma}$ is a function from superpositions of 
$n$ symbols words to superpositions of one symbol words 
(i.e. ``the local transition function", describing the way 
a cell interacts with its neghbours).\vspace{1mm}\\ 
Moreover $\delta$ must verify: \\ - the quiescent stability 
condition: $\big[\delta\ket{q^{n}})=\ket{q}\big]$;\\ - the 
no-nullity condition: $\forall w \in (q\Sigma)^{n},\, 
\big[\delta\ket{w}\neq\mathbf{0}\big].$
\end{Wdef}
By `successive' we mean that the number follow 
each other in unit step, i.e. the neighbourhood is an interval. 
In the literature these are sometimes referred to as simple
neighbourhoods, but it is trivial to simulate non-simple neighbourhoods automata with 
simple neighbourhoods automata. At this 
point we need not have a normalization condition such as 
$\forall w \in 
(q\Sigma)^{n},\,\big[||\delta\ket{w}||=1\big]$.
\noindent Configurations hold the basic states of an entire 
row of cells. As we will now formalize ours are finite but unbounded.
Note that fixed-sized periodic configurations\cite{VanDam}
as well as infinite configurations\cite{Schumacher} have
also been studied, leading to very different results and
proof methods (see Section \ref{sectionconclusion} for a
discussion). 
\begin{Def}[finite configurations, interval domains]~\\ A
\emph{(finite) 
configuration} $c$ of the quantum cellular automaton 
$\mathcal{A}=(\Sigma, q, N, \delta)$ is a function $c:
\mathbb{Z}\longrightarrow q\Sigma$, with $i\longmapsto
c(i)=c_i$, such that there exists a (possibly empty)
interval $I$ verifying $i\in I\Rightarrow c_i\in q\Sigma$
and $i\notin I\Rightarrow c_i=q$. Moreover we denote by 
$\textrm{idom}(c)$ the smallest such interval $I$, referred
to as the \emph{interval domain of $c$}. 
Then the interval $[k+\min(N),l+\max(N)]$ is denoted
$\textrm{extidom}(c)$ and referred to as the 
\emph{extended interval domain of $c$}. For the 
all-quiescent configuration $c=\ldots qqq \ldots$ we have 
$\textrm{idom}(c)=\emptyset$, and we let 
$\textrm{extidom}(c)=\emptyset$ also. Finally the set of 
all finite configurations is denoted $\mathcal{C}_f$, 
whilst the set of configurations having interval domains 
comprised within an interval $J$ is denoted 
$\mathcal{C}^J_f$.
\end{Def}
\begin{Def}[indexing conventions]~\label{indexing}\\ Given a 
configuration $c$ of the quantum cellular automaton 
$\mathcal{A}=(\Sigma, q, N, \delta)$, we denote by 
$c_{k\ldots l}$ the word $c_k\cdot\ldots\cdot c_l$ if 
$k\leq l$, and the empty word $\varepsilon$ otherwise. Thus
in either case $c_{k\ldots l}\in (q\Sigma)^*$. Moreover we
denote by $c_{i+N}$ the word 
$c_{i+\min(N)\ldots i+\max(N)}$, and $c_{i+\tilde{N}}$ the 
word $c_{i+\min(N)\ldots i+\max(N)-1}$. Therefore we have 
$c_{i+N}\in (q\Sigma)^n$ and  $c_{i+\tilde{N}}\in 
(q\Sigma)^{n-1}$, respectively.
\end{Def}
\noindent Whilst configurations hold the basic states of an 
entire row of cells, and hence denote the possible basic 
states of the entire LQCA, the global state of a LQCA may 
well turn out to be a superposition of these. The following 
definition works because $\mathcal{C}_f$ is a countably 
infinite set. 
\begin{Def}[superpositions of finite 
configurations]~\label{superp}\\ A \emph{superposition of 
configurations} of the quantum cellular automaton 
$\mathcal{A}=(\Sigma, q, N, \delta)$ is a normalized 
element of $\mathcal{H}_{\mathcal{C}_f}$, the Hilbert space 
of configurations.
\end{Def}
\begin{Def}[global evolution]~\label{global}\\ The 
\emph{global evolution} of the quantum cellular automaton 
$\mathcal{A}=(\Sigma, q, N, \delta)$ is the linear 
operation defined by linear extension of its action upon 
the canonical orthonormal basis, as follows:
\begin{align*}
\Delta:\mathcal{H}_{\mathcal{C}_f}&\rightarrow\mathcal{H}_{\mathcal{C}_f}\\ 
\ket{c}&\mapsto\Delta\ket{c}\\ 
\Delta\ket{c}=&\bigotimes_{i\in\mathbb{Z}}\delta\ket{c_{i+N}}
\end{align*}
\end{Def}
The postulates of quantum theory impose that the global 
evolution should be unitary. 
\begin{Def}[Unitarity]~\label{unitarity}\\ The \emph{global 
evolution} $\Delta$ of the quantum cellular automaton 
$\mathcal{A}=(\Sigma, q, N, \delta)$ is said to be 
\emph{(finite) unitary} if and only if 
$\{\Delta\ket{c}\,|\,c\in\mathcal{C}_f\}$ is an 
orthonormal basis of $\mathcal{H}_{\mathcal{C}_f}.$\\
\end{Def}
The next three lemmas are known facts in linear algebra which 
will prove useful later on. 
\begin{Lem}[Norm-preservedness, 
norm of rows]\label{lemnpfs}~\\ Let $\Delta: 
\mathcal{H}_S\rightarrow\mathcal{H}_S$ be a linear operator 
over a Hilbert space $\mathcal{H}_S$ having canonical 
orthonormal basis $\{\ket{c}\}_{c\in S}$. Suppose the 
following two conditions are fulfilled simultaneously:\\ 
-(i) $\forall c\in S,\,[||\Delta\ket{c}||=1]$;\\ -(ii) 
$\forall c,c'\in 
S,\,[\bra{c'}\Delta^{\dagger}\Delta\ket{c}\neq 
0\,\Leftrightarrow\,c=c']$;\\ Then $\forall r\in 
S,\,[0\leq||\Delta^{\dagger}\ket{r}||\leq1]$.\\
\end{Lem}
\textbf{Proof:} Conditions (i) and (ii) express the fact 
that $\Delta$ is norm-preserving. As a consequence  for all 
$r\in S$ we have 
$||\Delta^{\dagger}\ket{r}||=||\Delta\Delta^{\dagger}\ket{r}||$. 
Moreover we have by definition 
$||\Delta^{\dagger}\ket{r}||^2=|\bra{r}\Delta\Delta^{\dagger}\ket{r}|$. 
But the latter ($\bra{r}\Delta\Delta^{\dagger}\ket{r}$) is 
a projection of the former 
($\Delta\Delta^{\dagger}\ket{r}$) over a unit vector 
($\ket{r}$), and hence 
$|\bra{r}\Delta\Delta^{\dagger}\ket{r}|\leq||\Delta\Delta^{\dagger}\ket{r}||$. 
Therefore 
$||\Delta^{\dagger}\ket{r}||^2\leq||\Delta^{\dagger}\ket{r}||$ 
and so $||\Delta^{\dagger}\ket{r}||\leq 1$. \hfill$\Box$\\ 
As a corollary we have: \begin{Lem}[Norm-preservedness, 
finite spaces]\label{lemsumnorms}~\\ Let $\Delta: 
\mathcal{H}_S\rightarrow\mathcal{H}_S$ be a linear operator 
over a Hilbert space $\mathcal{H}_S$ having canonical 
orthonormal basis $\{\ket{c}\}_{c\in S}$. Let $T$ be a 
finite subset of $S$. Suppose the following three 
conditions are fulfilled simultaneously:\\ -(i) $\forall 
c\in S,\,[||\Delta\ket{c}||=1]$;\\ -(ii) $\forall c,c'\in 
S,\,[\bra{c'}\Delta^{\dagger}\Delta\ket{c}\neq 
0\,\Leftrightarrow\,c=c']$;\\ -(iii) $[\,\sum_{r \in 
T}||\Delta^{\dagger}\ket{r}||^2=|T|\,]$.\\ Then we have 
$\forall r\in T,\,[\,||\Delta^{\dagger}\ket{r}||=1\,]$.
\end{Lem}
\begin{Lem}[Unitarity from unit rows]~\\ Let $\Delta: 
\mathcal{H}_S\rightarrow\mathcal{H}_S$ be a linear operator 
over a Hilbert space $\mathcal{H}_S$ having canonical 
orthonormal basis $\{\ket{c}\}_{c\in S}$. Suppose the 
following three conditions are fulfilled simultaneously:\\ 
-(i) $\forall c\in S,\,[||\Delta\ket{c}||=1]$;\\ -(ii) 
$\forall c,c'\in 
S,\,[\bra{c'}\Delta^{\dagger}\Delta\ket{c}\neq 
0\,\Leftrightarrow\,c=c']$;\\ -(iii) $\forall r\in 
S,\,[||\Delta^{\dagger}\ket{r}||=1]$.\\ Then $\Delta$ is 
unitary.
\end{Lem}
\textbf{Proof:} Conditions (i) and (ii) express the fact 
that $\{\Delta\ket{c}\,|\,c\in S\}$ is an orthonormal set, 
i.e. that $\Delta$ is norm-preserving. Condition (iii) 
expresses the fact that for all $r\in S,\, 
\Delta^{\dagger}\ket{r}$ has unit norm. As a consequence 
$\Delta\Delta^{\dagger}\ket{r}$  has unit norm on the one 
hand, and 
$\bra{r}\Delta\Delta^{\dagger}\ket{r}=||\Delta^{\dagger}\ket{r}||=1$ 
on the other hand. Therefore 
$\alpha\Delta\Delta^{\dagger}\ket{r}=\ket{r}$, with 
$\alpha$ a root of unity. Since $\Delta\Delta^{\dagger}$ is 
positive, this $\alpha$ is just 1. Let 
$\Delta^{\dagger}\ket{r}=\sum \beta_c \ket{c}$. Then 
$\ket{r}=\sum \beta_c \Delta\ket{c}$, in other words each 
of the canonical orthonormal basis vectors $\ket{r}$ may be 
expressed as a linear combination of columns 
$\{\Delta\ket{c}\,|\,c\in S\}$. Therefore the columns form 
themselves an orthonormal basis.\hfill$\Box$\\
\noindent Next we will examine each of the following
conditions in 
turn:\\ -(i) the columns of $\Delta$ have unit norm 
(section \ref{sectionunitcols});\\ -(ii) the columns of 
$\Delta$ are orthogonal (section \ref{sectionorthocols});\\ 
-(iii) the rows of 
$\Delta$ have unit norm (section \ref{sectionrows}).

\section{Unit columns} \label{sectionunitcols}

The next two lemmas are simple facts from \cite{Durr}.
\begin{Lem}[Norm of a column]\label{lemnormcol}
Let $\Delta$ denote the \emph{global evolution} of the 
quantum cellular automaton $\mathcal{A}=(\Sigma, q, N, 
\delta)$. We have that $$\forall 
c\in\mathcal{C}_f,\quad\big[||\Delta\ket{c}||=\prod_{i\in\mathbb{Z}}||\delta\ket{c_{i+N}}||\big]$$
\end{Lem}
\textbf{Proof.} The norm of a tensor product of vectors is
the product of the norms of the vectors. \hfill $\Box$
\begin{Lem}[ Expressiveness of normalized $\mathbb{\delta}$]\label{lemexpressnormdelta}
Let $\Delta$ denote the global evolution of the quantum 
cellular automaton $\mathcal{A}=(\Sigma, q, N, \delta)$. 
Let $\Delta'$ denote the global evolution of the quantum 
cellular automaton $\mathcal{A}'=(\Sigma, q, N, \delta')$, 
with $\delta'$ such that $\forall 
w\in(q\Sigma)^{n},\,\big[\delta'\ket{w}=\delta\ket{w}/||\delta\ket{w}||\big].$ 
Suppose that the columns of $\Delta$ have unit norm. Then 
we have $\Delta=\Delta'$.
\end{Lem}
\textbf{Proof.} For all $c$ in $\mathcal{C}_f$ we have
\begin{align*}
\Delta\ket{c}=&\bigotimes_{i\in\mathbb{Z}}\delta\ket{c_{i+N}}=\bigotimes_{i\in\mathbb{Z}}||\delta\ket{c_{i+N}}||.\delta'\ket{c_{i+N}}\\ 
=&\prod_{i\in\mathbb{Z}}||\delta\ket{c_{i+N}}||.\bigotimes_{i\in\mathbb{Z}}\delta'\ket{c_{i+N}}\\ 
=&||\Delta\ket{c}||.\bigotimes_{i\in\mathbb{Z}}\delta'\ket{c_{i+N}}=\Delta'\ket{c} \qquad\Box
\end{align*}

\noindent Our approach is to change the actual definition of
LQCA as a consequence. 

\begin{Wdef}[LQCA]\label{oqca2}~\\ A linear quantum cellular 
automaton (LQCA) is a 4-tuple $\mathcal{A}=(\Sigma, q, N, 
\delta)$, where:\vspace{1mm}\\
- $\Sigma$ is a finite set of symbols (``the 
alphabet");\\
- $q$ is a symbol such that $q\notin\Sigma$ (``the 
quiescent symbol");\\
- $N$ is a set of n successive signed integers (``the 
neighboorhood");\\
- $\delta: \mathcal{H}_{(q\Sigma)^{n}}\rightarrow 
\mathcal{H}_{q\Sigma}$ is a function from superpositions of 
$n$ symbols words to superpositions of one symbol words 
(``the local transition function").\vspace{1mm}\\
Moreover $\delta$ must verify the following two 
properties:\\ - the quiescent stability condition: 
$\big[\delta\ket{q^{n}})=\ket{q}\big].$\\ - the 
normalization condition:\\ $\forall w\in 
(q\Sigma)^{n},\,\big[||\delta\ket{w}||=1\big].$\\
\end{Wdef}

\noindent This modified definition of LQCA  
choses to impose that $\forall w\in 
(q\Sigma)^{n},\,\big[||\delta\ket{w}||=1\big]$, i.e. that 
the local transition function $\delta$ is normalized. Then 
the fact that the columns of $\Delta$ have unit norm 
follows straight from lemma \ref{lemnormcol}. We are 
strongly justified to place this algebraic, almost 
syntactic restriction straight into the definition of LQCA 
because:\\ - The alternative is to have various 
non-normalized states compensating each other non-locally, 
which from a physical point of view is somewhat 
disturbing;\\ - It saves us from having to employ more 
elaborate techniques to check that columns have unit 
norms\cite{Durr}\cite{Hoyer}, such as applying least 
path algorithm to the associated de Bruijn graphs of the
quantum cellular automata etc. Although very elegant these tend
to render quantum cellular automata much more oblivious as a
model of computation;\\ - The modification made has absolutely no 
cost in terms expressiveness, as demonstrated in Lemma 
\ref{lemexpressnormdelta}.

\section{Orthogonality of columns}\label{sectionorthocols}

\noindent Having checked that the columns of the global
evolution matrix $\Delta$ have unit norm, we now turn to the
problem of deciding whether these columns are mutually
orhogonal. First we need a definition.
\begin{Def}[$A$-matrix]
Consider a linear quantum cellular automaton $\mathcal{A}=(\Sigma, q, N, \delta)$.
We call $A=\big[A^{\sigma}_{xy}\big]$
with $x,y\in(q\Sigma)^{n-1}$ and $\sigma\in(q\Sigma)$, the
matrix (tensor) such that $A_{xy}=\sum_\sigma
A^{\sigma}_{xy}\ket{\sigma}$ equals
$\delta\ket{w}$ if we have both $x=w_{1\ldots n-1}$ and
$y=w_{2\ldots n}$ for some $w\in(q\Sigma)^n$, otherwise it
is the null vector.
\end{Def}
For instance consider the sample rule:
$\delta\ket{000}=\ket{0}$, $\delta\ket{001}=\ket{1}$,
$\delta\ket{010}=\ket{1}$, $\delta\ket{011}=\ket{0}$,
$\delta\ket{100}=\ket{0}$, $\delta\ket{101}=\ket{1}$, 
$\delta\ket{110}=\ket{1}$, $\delta\ket{111}=\ket{0}$.\\
Then the $A$-matrix of the sample rule is:
\begin{align*}
\left(\begin{array}{cccc} \ket{0}    &\mathbf{0} &\ket{0}
&\mathbf{0}\\ \ket{1}    &\mathbf{0}    &\ket{1}
&\mathbf{0}\\ \mathbf{0}    &\ket{1}    &\mathbf{0}
&\ket{1}\\
\mathbf{0}    &\ket{0}    &\mathbf{0}   &\ket{0}
\end{array}\right).
\end{align*}
E.g. since $01$ and $00$ do not ``follow each other'' the
entry $\bra{00}A\ket{01}$ holds the null vector. On the
other hand $01$ and $10$ do overlap correctly to form the
neighbourhood $010$, for which $\delta\ket{010}=\ket{1}$,
hence $\bra{10}A\ket{01}=\ket{1}$.\\
How can we check that \emph{all} columns of some global
evolution are mutually orthogonal, when there is infinitely
many of them? Our next proposition is crucial in that respect, as
it shows why the problem which might seem to be of an
infinite (undecidable) nature is indeed of a finite
(decidable) nature. 
\begin{Proposition}[Finite columns
 checks]\label{finiteorthogcheck}~\\ \emph{This result 
refers to working definition \ref{oqca2}}.\\ Consider the 
global evolution $\Delta$ of a quantum cellular automaton 
$\mathcal{A}=(\Sigma, q, N, \delta)$. Let $s$ be equal to $ 
|q\Sigma|^{2n\!-\!2}-1$ and $I$ be the interval $[0,s]$. The 
columns $\{\Delta\ket{c}\,|\,c\in\mathcal{C}_f\}$ are 
orthogonal if and only if the columns 
$\{\Delta\ket{c}\,|\,c\in\mathcal{C}^I_f\}$ are orthogonal.
\end{Proposition}
(See appendix \ref{pffiniteorthogcheck} for a detailed
proof.)\\
\noindent We now give our algebraic condition upon $\delta$
ensuring that the columns of $\Delta$ are mutually orthogonal. 
\begin{Proposition}[Column test]\label{algdls}~\\ 
\emph{This result refers to working definition 
\ref{oqca2}}.\\ 
Consider the global evolution $\Delta$ of a 
quantum cellular automaton $\mathcal{A}=(\Sigma, q, N, 
\delta)$. Let $s= |q\Sigma|^{2n\!-\!2}-1$. The columns 
$\{\Delta\ket{c}\,|\,c\in\mathcal{C}_f\}$ are orthogonal if 
and only if $\forall x,x'\in(q\Sigma)^{n\!-\!1}$:
\begin{equation}
\left[ 
\bra{xx'}M^s\ket{q^{n\!-\!1}q^{n\!-\!1}}\bra{q^{n\!-\!1}q^{n\!-\!1}}M^s\ket{xx'}\neq 
0 \Leftrightarrow (x=x')\right]  \label{algorthogeq}
\end{equation}
with $M=[M_{xx',yy'}],  M_{xx',yy'}=\big|\sum_\sigma 
A^{\sigma*}_{x'y'} A^{\sigma}_{xy}\big|^2$, $A$ the 
$A$-tensor of the LQCA.
\end{Proposition}
(See appendix \ref{pfalgdls} for a detailed
proof.)\\
\noindent We believe that the obtention of
algebraic conditions constituted a necessary step in order
to be able to take further the analysis of this
model. The algebraic proofs of these conditions give 
them a physical meaning which shortcuts the 
graph-theoretical detour, which is particularly useful since
quantum theory people tend to reason in terms of linear
algebra rather than graph theory. They also master the
corresponding numerical tools better, Proposition
\ref{algdls} makes it easy to check for column
orthonormality through any software tool which 
does matrix multiplication.\\
Still there remains some space for improvement, for instance
because Proposition 
\ref{algdls} is phrased in terms of the somewhat bizarre 
$A$-tensor of $\mathcal{A}$, rather than just $\delta$. 
Fortunately this first point can be fixed using the quantum 
equivalent of a simplifying classical 
result\cite{Ibarra}\cite{Pedersen}\cite{Boykett}: 
\begin{Lem}[Expressivity of size two 
$\delta$]\label{lempedersen}~\\ \emph{This result refers to 
working definition \ref{oqca2}}.\\ Consider a linear quantum 
cellular automaton $\mathcal{A'}=(\Sigma, q, N, \delta')$ 
and its global evolution $\Delta'$. One can always 
construct a linear quantum cellular automaton 
$\mathcal{A}=(\Sigma^{n-1}, q^{n-1}, \{0,\!1\}, \delta)$ 
such that its global evolution $\Delta$ equals $\Delta'$. 
Then the $A$-tensor of $\mathcal{A}$ is just 
$\delta=\big[\delta^{\sigma}_{xy}\big]$, with $x,y, 
\sigma\in\Sigma^{n-1}$.
\end{Lem}
\textbf{Proof.} We let, for all $x,y \in\Sigma^{n-1}$
\begin{equation*}
\delta\ket{x_1\ldots x_{n\!-\!1}y_1\ldots 
y_{n\!-\!1}}\equiv\bigotimes_{i\in[1,n\!-\!1]} 
\delta'\ket{x_i\ldots x_{n\!-\!1}y_1\ldots y_i}.
\end{equation*}
The rest follows from the definitions.\hfill$\Box$\\ 
Note that the groups of $n-1$ cells constructed in 
this lemma bear some resemblance with the reduced 
neighbourhoods $\tilde{N}$ constructed by D\"urr et al. -- 
except they do not overlap, which consequently saves us 
from using De Buijn graphs and the like. This suggests that in
spite of its apparent simplicity this trick is probably just the right 
way to enumerate/construct LQCA.\\ We change the definition
of LQCA as a consequence. 
\begin{Wdef}[LQCA]\label{oqca3}~\\ A linear quantum cellular 
automaton (LQCA) is a 3-tuple $\mathcal{A}=(\Sigma, q, 
\delta)$, where:\vspace{1mm}\\
- $\Sigma$ is a finite set of symbols (``the 
alphabet");\\
- $q$ is a symbol such that $q\notin\Sigma$ (``the 
quiescent symbol");\\
- $\delta: \mathcal{H}_{(q\Sigma)^{2}}\rightarrow 
\mathcal{H}_{q\Sigma}$ is a function from superpositions of 
$2$ symbols words to superpositions of one symbol words 
(``the local transition function").\vspace{1mm}\\
Moreover $\delta$ must verify the following two 
properties:\\ - the quiescent stability condition: 
$\big[\delta\ket{qq})=\ket{q}\big].$\\ - the normalization 
condition:\\ $\forall w\in 
(q\Sigma)^{2},\,\big[||\delta\ket{w}||=1\big].$\\
\end{Wdef}
This modified definition of LQCA choses to 
impose neighbourhoods of size two on top of the 
normalization condition. For these linear quantum cellular 
automata we define $\Delta$ as usual with $N=\{0,1\}$. 
Again we are strongly justified to place this easy 
restriction straight into the definition of LQCA for it 
simplifies and makes more intuitive the decision 
procedure induced by Proposition \ref{algdls}, (the 
corresponding simplifications are shown in Corollary 
\ref{summary}). The modification made comes at absolutely 
no cost in terms expressiveness, as demonstrated in Lemma 
\ref{lempedersen}.\\
These successive two modifications we have made to our model
will be even more asserted by the important simplification
they bring to the problem of determining whether a LQCA
has unit rows.
 
\section{Unit rows}\label{sectionrows}

\noindent Having checked that the columns of the global
evolution 
matrix $\Delta$ are orthonormal, we now turn to the problem 
of deciding whether its rows have unit norm.
\begin{Proposition}[row norm as matrix 
product]\label{rowasmat}~\\ \emph{This result refers to 
working definition \ref{oqca3}.}\\ Consider a quantum 
cellular automaton $\mathcal{A}=(\Sigma, q, \delta)$ whose 
global evolution $\Delta$ has orthonormal columns. The 
squared norm of any row $r$ is given by 
$$||\Delta^{\dagger}\ket{r}||^2=\lim_{h\rightarrow\infty}\bra{q}{N^{(q)}}^{h}\big(\prod_{i\in 
k\ldots l}  N^{(r_{i})}\big){N^{(q)}}^{h} \ket{q}$$ where 
$\{N^{(\sigma)}\}_{\sigma\in q\Sigma}$ is the set of 
matrices such that $N^{(\sigma)}=[N^{(\sigma)}_{x,y}], 
N^{(\sigma)}_{x,y}=|\bra{\sigma}\delta\ket{xy}|^2$.
\end{Proposition}
(See appendix \ref{pfrowasmat} for a detailed proof.)\\
The following proposition takes advantage of the successive 
restrictions made in definitions \ref{oqca2} and 
\ref{oqca3} to bring about a crucial simplification -- 
which makes obsolete a good half of the procedure 
described in \cite{Durr2}.

\begin{Proposition}[Middle 
segment]\label{localunitcheck}~\\ \emph{This result refers 
to working definition \ref{oqca3}.}\\ Consider a quantum 
cellular automaton $\mathcal{A}=(\Sigma, q, \delta)$ whose 
global evolution $\Delta$ has orthonormal columns. The rows 
$\{\Delta^{\dagger}\ket{r}\,|\,r\in\mathcal{C}_f\}$ have 
unit norm if and only if $$\lim_{h\rightarrow\infty}\bra{q} 
{N}^{h}O{N}^{h}\ket{q}=|q\Sigma|$$ with $O=[1_{xy}]$ the 
matrix with only ones, and $N=[N_{x,y}], 
N_{x,y}=|\bra{q}\delta\ket{xy}|^2$.
\end{Proposition}
(See appendix \ref{pflocalunitcheck} for a detailed proof.)\\
Note that $\lim_{h\rightarrow\infty}\bra{q} 
{N}^{h}O{N}^{h}\ket{q}= \sum_{r\in\mathcal{C}^{[0,0]}_f} 
||\Delta^{\dagger}\ket{r}||^2$. Hence we have the following 
insightful corollary, which comes as the direct analogue of 
our Proposition \ref{finiteorthogcheck}. Curiously however it 
will not contribute to our final result. 
\begin{Proposition}[Finite 
rows check]\label{finiterowcheck}~\\ \emph{This result 
refers to working definition \ref{oqca3}.}\\ Consider a 
quantum cellular automaton $\mathcal{A}=(\Sigma, q, 
\delta)$ whose global evolution $\Delta$ has orthonormal 
columns. The rows 
$\{\Delta^{\dagger}\ket{r}\,|\,r\in\mathcal{C}_f\}$ have 
unit norm if and only if the rows 
$\{\Delta^{\dagger}\ket{r}\,|\,r\in\mathcal{C}^{[0,0]}_f\}$ 
have unit norm.
\end{Proposition}

\noindent Both results deepen our understanding of the
algebraic structure of unitary linear quantum cellular 
automata. Proposition \ref{localunitcheck} offers an 
important simplification along the way to determining 
whether the global evolution $\Delta$ has unit rows, 
reducing this to the evaluation of $\ora{l}.O\ora{r}$.\\
The difficult problem we are left with is that of 
evaluating the so-called `border vectors' \cite{Durr2} 
$\ora{r}=\lim_{h\rightarrow\infty}{N}^{h}\ket{q}$ and 
$\ora{l}=\lim_{h\rightarrow\infty}{N^{\dagger}}^{h}\ket{q}$. The 
following proposition will characterize them uniquely and 
algebraically. We let $\leq$ denotes the following partial 
order upon $m\times n$ matrices: $$ M\leq N 
\Longleftrightarrow \forall i,j \quad M_{ij}\leq N_{ij}.$$ 
Column vectors are seen as $m\times 1$ matrices for that 
matter. 
\begin{Proposition}[Border 
vectors]\label{minbordervecs}~\\ \emph{This result refers 
to working definition \ref{oqca3}.}\\ Consider a quantum 
cellular automaton $\mathcal{A}=(\Sigma, q, \delta)$ whose 
global evolution $\Delta$ has orthonormal columns. The 
vectors 
$$\ora{r}=\lim_{h\rightarrow\infty}{N}^{h}\ket{q}\quad\textrm{and}\quad\ora{l}=\lim_{h\rightarrow\infty}{{N}^{\dagger}}^{h}\ket{q}$$ 
with $N=[N_{x,y}], N_{x,y}=|\bra{q}\delta\ket{xy}|^2$, have 
only finite entries. They verify
\begin{align*}
\ora{r}&=\min_{\leq}\{v\;|\;\; \mathbf{0}\leq v\;\wedge\; 
Nv=v\;\wedge\;v_q=1\}\\ 
\textrm{and}\quad\ora{l}&=\min_{\leq}\{v\;|\;\; 
\mathbf{0}\leq v\;\wedge\; vN=v\;\wedge\;v_q=1\}.
\end{align*}
Moreover the following extra conditions hold:\\ (i) 
$\ora{l}.\ora{r}=1;$\\ (ii)  $(\sum_i \ora{l}_i)(\sum_i 
\ora{r}_i)\leq |q\Sigma|$.\\ (iii) $\forall x\in\Sigma,\; 
[\ora{l}_x=0\;\vee\;\ora{r}_x=0];$\\ (iv) $\forall 
x,y\in\Sigma,\; [N_{xy}\neq 
0\Rightarrow\ora{l}_x=0\;\vee\;\ora{l}_y=0]$;\\ Inequality 
(ii) is saturated if and only if $\Delta$ has unit rows.
\end{Proposition}
(See appendix \ref{pfminbordervecs} for a detailed proof.)
\noindent This last proposition is highly informative, and in 
\emph{most cases} will provide us with an effective way to 
compute these border vectors, through a spectral 
decomposition of $N$. When the eigenvalue $1$ is 
degenerate, however, it is unclear to the author whether 
there exists a definite procedure to performing the 
minimization.\\ 

\noindent Taking a step back from the mathematics, we may 
wonder where these limits $N^h$, $h\rightarrow\infty$ come 
from `physically'. Say we wish to calculate the norm of a 
row $\Delta^{\dagger}\ket{r}$, corresponding to a 
configuration $r$ with interval domain $I$. Then at some 
point we need to sum over all the antecedent configurations 
of $r$ (cf. the beginning of the proof of Proposition 
\ref{rowasmat}). Our issue now arises from the fact that 
although the antecedents of $r$ are in $\mathcal{C}_f$, 
there may be an infinite number of them, and hence they may 
not be able to restrict to some interval $J$. A good 
example of this is provided in \cite{Durr2}, which we now 
reproduce for convenience:
\begin{Rk}
Consider $\textsc{Qflip}=({p}, q, \delta)$ with 
$\delta\ket{qq}=\ket{q}$, 
$\delta\ket{qp}=\ket{q}+\ket{p}/\sqrt{2}$, 
$\delta\ket{pq}=\ket{p}$, 
$\delta\ket{pp}=\ket{q}-\ket{p}/\sqrt{2}$. Its 
corresponding global evolution $\Delta$ is unitary. Let 
$c^n=\ldots qqp^nqq\ldots$, with the rightmost $p$ in 
position $0$. Then 
$\bra{c^1}\Delta\ket{c^n}=(1/\sqrt{2})^n$.
\end{Rk}
Faced with such situations our temptation is again to 
further restrict the definition of LQCA, with a view to 
eliminate such scenarios. For instance we may want to add 
to the Working definition \ref{oqca3} a \emph{full 
stability condition} asking that:\\ $\forall w\in 
\Sigma^{2},\,\big[\bra{q}\delta\ket{w}=0\big].$\\
This extra `full stability condition' would ensure that the 
antecedents $c$ of a configuration $r$ have an interval 
domain no greater that the extended interval domain of $r$. 
\begin{Lem}[Full stability unit 
rows]\label{localunitcheck2}~\\ \emph{This result refers to 
working definition \ref{oqca3}.}\\ Consider a quantum 
cellular automaton $\mathcal{A}=(\Sigma, q, \delta)$ whose 
global evolution $\Delta$ has orthonormal columns. Suppose 
the full stability condition $\forall w\in 
\Sigma^{2},\,\big[\bra{q}\delta\ket{w}=0\big].$ is 
verified. Then the rows 
$\{\Delta^{\dagger}\ket{r}\,|\,r\in\mathcal{C}_b\}$ have 
unit norm if and only if
\begin{equation}
\bra{q} {N}{O}{N}\ket{q}=|q\Sigma| \label{algnormeq2}
\end{equation}
with ${O}=[1_{xy}]$ the matrix with only ones, and 
$N=[N_{x,y}], N_{x,y}=|\bra{q}\delta\ket{xy}|^2$.
\end{Lem}
\textbf{Proof.}  First note that under this condition, one 
has $\forall x\in\Sigma$
\begin{align*}
&\bra{x}N^2\ket{q}=\sum_{\sigma\in 
q\Sigma}N_{x\sigma}N_{\sigma q}=\sum_{\sigma\in 
q\Sigma}|\bra{q}\delta\ket{x\sigma}|^2|\bra{q}\delta\ket{\sigma 
q}|^2\\ 
&=|\bra{q}\delta\ket{xq}|^2|\bra{q}\delta\ket{qq}|^2\quad\textrm{by 
the full stability condition}\\ 
&=|\bra{q}\delta\ket{xq}|^2\quad\textrm{by the quiescent 
stability condition}\\ &=\bra{x}N\ket{q}.
\end{align*}
Moreover we have
$\bra{q}N^2\ket{q}=\sum_{\sigma\in 
q\Sigma}N_{q\sigma}N_{\sigma q}$, which is equal to
$N_{qq}N_{qq} = 1= \bra{q}N\ket{q}$ using the fact that 
$\forall x\in\Sigma\; [N_{qx}N_{xq}=0]$. The contrary would 
imply $\bra{\ldots qq \ldots}\Delta\ket{\ldots qqxqq 
\ldots}\neq 0$, which is impossible since 
$\bra{q}\delta\ket{qq}=1$ and $\Delta$ is 
norm-preserving.\\ As a consequence we have 
$N^2\ket{q}=N\ket{q}$. Symmetrically so for 
$\bra{q}N^2=\bra{q}N$. The rest follows from Proposition 
\ref{localunitcheck}.\hfill$\Box$\\
Unfortunately it is not clear whether this extra 
restriction of LQCA implies a loss of expressiveness. In 
fact may even disqualifies some valid reversible cellular 
automata from being quantum cellular automata. We leave the 
question of a simulation as an open problem. \begin{Open} 
Consider a linear quantum cellular automaton 
$\mathcal{A}=(\Sigma, q, \delta)$ following to definition 
\ref{oqca3} and its global evolution $\Delta$. Is it always 
possible to construct a linear quantum cellular automaton 
$\mathcal{A'}=(\Sigma', q', \delta')$ such that $\forall 
w\in \Sigma^{2},\,\big[\bra{q}\delta\ket{w}=0\big]$ and its 
global evolution $\Delta'$ equals $\Delta$? (We may want to 
admit loser notions of efficient simulation.)
\end{Open}
In fact one may wonder whether there may not be a way to
make any norm-preserving LQCA (unit, orthogonal columns)
into a unitary LQCA exhibiting the full stability condition.
For instance: 
\begin{Rk}
Consider $\textsc{Xor}=(\{0,1\}, 0, \delta)$ with 
$\delta\ket{00}=\ket{0}$, $\delta\ket{01}=\ket{1}$, 
$\delta\ket{10}=\ket{1}$, $\delta\ket{11}=\ket{0}$. Its  
corresponding global evolution $\Delta$ is injective in the 
space of finite configurations (i.e. it is norm-preserving, 
it has orthonormal columns). But the rule
$\delta\ket{11}=\ket{0}$ breaks the full stability
condition. Moreover it is not surjective (i.e. its rows are
not unit norm) since $\ldots 00100\ldots$ does not have any antecedent.
\end{Rk}
is suitably 'fixed' as in Remark \ref{rkfix}.

\section{Conclusion}\label{sectionconclusion}

The following theorem is a synthesis our results. 
\begin{Th}[Summary]\label{summary}~\\ \emph{This result 
refers to working definition \ref{oqca3}.}\\ Consider the 
global evolution $\Delta$ of a quantum cellular automaton 
$\mathcal{A}=(\Sigma, q, \delta)$. Let $s= 
|q\Sigma|^{2}-1$. $\Delta$ is unitary if and only if 
$\forall x,x'\in(q\Sigma)$:\\ \begin{align*} &\left[ 
\bra{xx'}M^s\ket{qq}\bra{qq}M^s\ket{xx'}\neq 0 
\Leftrightarrow (x=x')\right]\\ &\textrm{and}\quad (\sum_i 
\ora{l}_i)(\sum_i \ora{r}_i)=|q\Sigma|\\ 
\textrm{with}\;&\ora{r}=\min_{\leq}\{v\;|\;\; \mathbf{0}\leq v\;\wedge\; 
Nv=v\;\wedge\;v_q=1\};\\ 
&\ora{l}=\min_{\leq}\{v\;|\;\; 
\mathbf{0}\leq v\;\wedge\; vN=v\;\wedge\;v_q=1\};\\ 
&M=[M_{xx',yy'}],\;M_{xx',yy'}=\big|\bra{x'y'}\delta^{\dagger}\delta\ket{xy}\big|^2;\\ 
&N=[N_{x,y}],\;N_{x,y}=|\bra{q}\delta\ket{xy}|^2.
\end{align*}
\end{Th}
We have definitely gone a long way towards the 
simplification and the algebraization of unitarity criteria 
for one dimensional quantum cellular automata as defined in 
\cite{Watrous}, \cite{Durr}, \cite{Durr2}. Note that these 
last two papers do not contain any such synthetic algebraic criteria. 
Instead they provide several pages long decision 
procedures, which have many twists and bends.\\ 
Certainly we have 
not gone as far as to reduce LQCA to partitioned quantum 
cellular automata (PQCA) or quantum cellular automata with 
Margolus neighbourhood (MQCA) etc., i.e. our global 
evolution cannot, in general, be decomposed into the 
application of one small unitary operator homogeneously 
across space.\\ Other models of quantum cellular automata 
do admit such reductions; the problem of deciding unitarity 
becomes incomparably easier, or even trivial by 
construction \cite{Schumacher}. Such crucial differences 
seem to arise when one considers different spaces of 
configurations, e.g. finite periodic \cite{VanDam}, because 
these constrain reversibility to be structural, i.e. an 
essentially local matter. With the space of finite 
configurations $\mathcal{C}_f$ reversibility becomes a 
global matter, even though the global evolution is defined 
locally. Analogues of this are well-known in the classical 
realm already:
\begin{Rk}\label{rkfix}
Consider $\textsc{Xor'}=(\{q,0,1\}, q, \delta)$ with 
$\delta\ket{00}=\ket{0}$, $\delta\ket{01}=\ket{1}$, 
$\delta\ket{10}=\ket{1}$, $\delta\ket{11}=\ket{0}$,
$\delta\ket{q0}=\ket{q}$, $\delta\ket{q1}=\ket{q}$,
$\delta\ket{0q}=\ket{0}$, $\delta\ket{1q}=\ket{1}$. Its 
corresponding global evolution $\Delta$ is bijective in the 
space of finite configurations (i.e. it is unitary), 
but the global evolution cannot be reversed by a cellular automata. 
Moreover $\Delta$ is not reversible in the space of infinite 
configurations ($\ldots 00 \ldots$ may have antecedents 
either $\ldots 00 \ldots$ or $\ldots 11\ldots$).
\end{Rk}
One may argue that PQCA models are enough, since they 
simulate the quantum Turing machine. If we are interested 
in 'intrinsic universality' however, and want to consider
the above example as 'physical' then these are
not be enough. A local transition $\delta$ which is 
unitary in the space of infinite configurations is 
also unitary in the space if finite configurations, 
but the reciprocal statement is not true. In this sense 
there seems to be a loss of expressiveness when restricting 
to PQCA models.\\ 
Note that this research constitutes a first step in the quest for the 
identification of the one particular unitary LQCA known to 
simulate all other LQCA efficiently. Obviously as we 
restrict the number of one dimensional quantum cellular 
automata to be considered to a well-behaved subclass, the 
quest for such an `intrinsically universal'  one 
dimensional quantum cellular automata must become easier. 
Moreover our approach may eventually open this 
(well-advertised in the literature) open problem to an 
algebraic analysis.\\ Note also that it is undecidable 
whether local transition function $\delta$ induces a 
unitary global evolution $\Delta$ as soon as the quantum 
cellular automata is two dimensional. But again this does 
not have to be the end of it: nothing prevents that there 
should be a canonical definition of two dimensional quantum 
cellular automata, easily checked for unitarity, and yet 
capable of expressing the exact same set of global 
evolutions. For this purpose algebraic criteria ought to be 
easier to generalize to higher dimensions.

\section*{Acknowledgments} P.J.A  would like to thank Shu 
Yan Chan, Christoph D\"urr, Simon Perdrix, and Estrella 
Sicardi for a number of insightful conversations, and 
acknowledges the direct help of Thierry Gallay in proving 
proposition \ref{minbordervecs}. This research was 
initiated as an Assistant Professor in Montevideo, Uruguay, 
at the Institute of Physics of the Engineering Faculty and 
concluded as a CNRS post-doc in Grenoble, France, at the 
Leibniz Laboratory. Towards both institutions the author is 
greatly indebted of a very pleasant stay. Many thanks in 
particular to Gonzalo Abal, Arturo Lezama, Alejandro 
Romanelli, Ricardo Siri and above all Philippe Jorrand for 
offering me these excellent working conditions.

\appendix
\section{Proofs of Propositions \ref{finiteorthogcheck}--\ref{minbordervecs}}

\subsection{Proof of Proposition \ref{finiteorthogcheck}.} 
\label{pffiniteorthogcheck}

Consider $c\neq c'$ such
that $\Delta\ket{c}$ is not 
orthogonal to $\Delta\ket{c'}$. Since there are 
$|q\Sigma|^{2n\!-\!2}$ possible pairs of $(q\Sigma)^{n-1}$ 
subwords, there must exist $k,l\in \mathbb{Z}$ with $0\leq 
l-k \leq s$ such that $c_{k\ldots k+n-2}=c_{l\ldots l+n-2}$ 
and $c'_{k\ldots k+n-2}=c'_{l\ldots l+n-2}$. We now 
construct $d=c_{\infty\ldots k+n-2}\cdot c_{l+n-1\ldots 
\infty}$ and $d'=c'_{\infty\ldots k+n-2}\cdot 
c'_{l+n-1\ldots \infty}$, and notice that
\begin{align*}
\big(\Delta\ket{d},\Delta\ket{d'}\big)&=\!\!\!\!\!\!\prod_{i\notin]k+n-2,l+n-1[}\!\!\!\!\!\big(\delta\ket{c'_{i+N}},\delta\ket{c_{i+N}}\big)\neq 
0\\ \textrm{since}\quad 
\big(\Delta\ket{c},\Delta\ket{c'}\big)&=\prod_i 
\big(\delta\ket{c'_{i+N}},\delta\ket{c_{i+N}}\big)\neq 0.
\end{align*}
By iterating this procedure the right number of times we 
obtain, from any $c\neq c'$ such that $\Delta(c)$ is not 
orthogonal to $\Delta(c')$, a pair $d\neq d'$ such that 
$\Delta(d)$ is not orthogonal to $\Delta(d')$ and with both 
$\text{idom}(d)$ and $\text{idom}(d')$ included in an 
interval of size less or equal to $s$. By shift-invariance 
it is therefore sufficient to check that the columns 
$\{\Delta\ket{c}\,|\,c\in\mathcal{C}^I_f\wedge\text{idom}(c)\leq 
s\}$ are pairwise orthogonal. \hfill$\Box$

\subsection{Proof of Proposotion \ref{algdls}.}
\label{pfalgdls}
Let $I$ stand for the interval $[0,s]$ and $J$ stand for 
the interval $[-s,s]$.\\ $\mathbb{[\Rightarrow]}\;$ 
Corollary \ref{finiteorthogcheck} implies that the columns 
$\{\Delta\ket{c}\,|\,c\in\mathcal{C}_f\}$ are orthogonal if 
and only if we have $\forall c,\, c'\in 
\mathcal{C}^J_f\quad 
|\big(\Delta\ket{c},\Delta\ket{c'}\big)|^2$ strictly 
positive when $c$ equals $c'$ and zero otherwise. :
\begin{align}
&\Leftrightarrow \!\!\!\!\!\!\prod_{i\in 
\textrm{extidom}(J)}\!\!\!\! (\delta\ket{ 
c_{i+N}})^{\dagger}(\delta\ket{ 
c'_{i+N}})\;\mathop{>}_{c=c'}\!\Big/\!\!\!\mathop{=}_{c\neq 
c'}\!0\nonumber\\ &\Leftrightarrow \!\!\!\!\!\!\prod_{i\in 
\textrm{extidom}(J)} \!\!\!\!(A\ket{c_{i+\tilde{N}}, 
c_{i+1+\tilde{N}}})^{\dagger} (A\ket{c'_{i+\tilde{N}}, 
c'_{i+1+\tilde{N}}})\mathop{>}_{c=c'}\!\Big/\!\!\!\mathop{=}_{c\neq 
c'}\!0\nonumber\\ &\Leftrightarrow \!\!\!\!\!\!\prod_{i\in
\textrm{extidom}(J)}\!\!\!\! \bra{c_{i+1+\tilde{N}} 
c'_{i+1+\tilde{N}}}M\ket{c_{i+\tilde{N}} 
c'_{i+\tilde{N}}}\mathop{>}_{c=c'}\!\Big/\!\!\!\mathop{=}_{c\neq 
c'}\!0\nonumber\\ &\Leftrightarrow \!\!\!\!\!\!\prod_{i\in
\textrm{extidom}(J)}\!\!\!\! M_{c_{i+1+\tilde{N}}
c'_{i+1+\tilde{N} , c_{i+\tilde{N}} c'_{i+\tilde{N}}} 
}\;\mathop{>}_{c=c'}\!\Big/\!\!\!\mathop{=}_{c\neq 
c'}\!0\nonumber
\end{align}
Summing this last equation over those configurations $c,c' 
\in\mathcal{C}^J_f$ which verify $c_{0+\tilde{N}}=x, 
c'_{0+\tilde{N}}=x'$ yields, for any fixed 
$x,x'\in\Sigma^{n-1}$:
\begin{align*}
\bra{q^{n\!-\!1}q^{n\!-\!1}} M^s\ket{xx'}\bra{xx'} M^s 
\ket{q^{n\!-\!1}q^{n\!-\!1}}\;\mathop{>}_{x=x'}\!\Big/\!\!\!\mathop{=}_{x\neq 
x'}\!0
\end{align*}
$\mathbb{[\Leftarrow]}\;$ Consider two configurations 
$d,d'\in\mathcal{C}^I_f$. By merely shifting to the center 
any difference between $d$ and $d'$ one can always 
construct $c,c'\in\mathcal{C}^J_f$ such that 
$c_{0+\tilde{N}}=x, c'_{0+\tilde{N}}=x'$, and 
$[x=x'\quad\textrm{if and only if}\quad d=d']$. Now 
consider condition (\ref{algorthogeq}):
\begin{align*}
&M_{q^{n\!-\!1}q^{n\!-\!1}, \ldots}\prod_i 
M_{c_{i+\tilde{N}} c'_{i+\tilde{N}}, 
c_{i+1+\tilde{N}}c'_{i+1+\tilde{N}}} M_{\ldots,xx'}\\ \cdot 
& M_{xx',\ldots}\prod_i M_{c_{i+\tilde{N}} 
c'_{i+\tilde{N}}, c_{i+1+\tilde{N}}c'_{i+1+\tilde{N}}} M_{ 
\ldots, 
q^{n\!-\!1}q^{n\!-\!1}}\;\mathop{>}_{x=x'}\!\Big/\!\!\!\mathop{=}_{x\neq 
x'}\!0\\ &\Leftrightarrow \!\!\!\!\!\!\prod_{i\in 
\textrm{extidom}(J)}\!\!\!\! (\delta\ket{ 
c_{i+N}})^{\dagger}(\delta\ket{ 
c'_{i+N}})\;\mathop{>}_{c=c'}\!\Big/\!\!\!\mathop{=}_{c\neq 
c'}\!0\nonumber\\ &\Leftrightarrow \!\!\!\!\!\!\prod_{i\in 
\textrm{extidom}(I)}\!\!\!\! (\delta\ket{ 
d_{i+N}})^{\dagger}(\delta\ket{ 
d'_{i+N}})\;\mathop{>}_{d=d'}\!\Big/\!\!\!\mathop{=}_{d\neq 
d'}\!0\nonumber\\
\end{align*}
In other words the columns 
$\{\Delta\ket{d}\,|\,d\in\mathcal{C}^I_f\}$ are orthogonal, 
and so Corollary \ref{finiteorthogcheck} implies that the 
columns $\{\Delta\ket{c}\,|\,c\in\mathcal{C}_f\}$ are 
orthogonal.\hfill$\Box$

\subsection{Proof of Proposition \ref{rowasmat}} \label{pfrowasmat}
\begin{align*}
\Delta&=\sum_{r,c\in\mathcal{C}_f} \ket{r}\bra{r} \Delta
\ket{c}\bra{c} \\ &=\sum_{r,c\in\mathcal{C}_f} 
\ket{r}\bra{r} \bigotimes_{i\in\mathbb{Z}} 
\delta\ket{c_{i}c_{i+1}}\bra{c}\\
\bra{r}\Delta&=\sum_{c\in\mathcal{C}_f} 
\prod_{i\in\mathbb{Z}} \bra{r_i} 
\delta\ket{c_{i}c_{i+1}}\bra{c}\\
\bra{r}\Delta\Delta^{\dagger}\ket{r}&=\sum_{c\in\mathcal{C}_f} 
\prod_{i\in\mathbb{Z}} | \bra{r_i} 
\delta\ket{c_{i}c_{i+1}}|^2\\ 
||\Delta^{\dagger}\ket{r}||^2&= \sum_{c\in\mathcal{C}_f} 
\prod_{i\in\mathbb{Z}} N^{(r_{i})}_{c_{i}c_{i+1}}
\end{align*}
Say $r$ has an interval domain included in $I=[k, l]$. Then 
$||\Delta^{\dagger}\ket{r}||^2$ equals
\begin{align*}
\sum_{c\in\mathcal{C}_f} \prod_{i\in -\infty\ldots k-1} 
N^{(q)}_{c_{i}c_{i+1}} \prod_{i\in k\ldots l} 
N^{(r_{i})}_{c_{i}c_{i+1}} \prod_{i\in l+1\ldots \infty} 
N^{(q)}_{c_{i}c_{i+1}}
\end{align*}
If we now restrict our sum to antecedents whose interval 
domain is included in $J=[k-h, l+h+1]$, each term in the 
sum takes the form
\begin{align*}
&N^{(q)}_{qq}\cdots 
N^{(q)}_{qc_{k-h}}\!\!\!\!\!\!\!\!\!\cdots 
N^{(q)}_{c_{k-1}c_{k}}(\!\!\prod_{i\in k\ldots l}\!\!\!\! 
N^{(r_{i})}_{c_{i}c_{i+1}}) N^{(q)}_{c_{l+1}c_{l+2}} 
\!\!\!\!\!\!\!\!\cdots N^{(q)}_{c_{l+h+1}q}\!\!\!\!\cdots 
N^{(q)}_{qq} \\ &=N^{(q)}_{qc_{k-h}}\!\!\cdots 
N^{(q)}_{c_{k-1}c_{k}}(\prod_{i\in k\ldots l} 
N^{(r_{i})}_{c_{i}c_{i+1}}) N^{(q)}_{c_{l+1}c_{l+2}} 
\!\!\cdots N^{(q)}_{c_{l+h+1}q}\\ &\textrm{using } 
N^{(q)}_{qq}=1. \;\; \textrm{Performing the sum yields}\\ 
&\bra{q} \big(\prod_{k-1-h\ldots k-1} 
N^{(q)}\big)\big(\prod_{i\in k\ldots l} 
N^{(r_{i})}\big)\big(\prod_{l+1\ldots l+h+1}  N^{(q)}\big) 
\ket{q}\\ &=\bra{q} {N^{(q)}}^{h}\big(\prod_{i\in k\ldots 
l} N^{(r_{i})}\big){N^{(q)}}^{h} \ket{q}
\end{align*}
The limit of this expression as $h$ tends to infinity 
remains finite. Indeed, let us suppose that the contrary is 
true. This implies that the row $\Delta^{\dagger}\ket{r}$ 
has infinite norm. But by lemma \ref{lemnpfs} this 
contradicts the fact that $\Delta$ is assumed to be 
norm-preserving.\hfill$\Box$

\subsection{Proof of Proposition \ref{localunitcheck}}
\label{pflocalunitcheck}

By lemma \ref{lemsumnorms},
requiring
that the rows be of unit norm of rows is equivalent to the 
condition that for all $t$ and $J=[-t,t]$, 
$$\sum_{r\in\mathcal{C}^J_f} 
||\Delta^{\dagger}\ket{r}||^2=|q\Sigma|^{2t+1}.$$ Using 
proposition \ref{rowasmat} the equation becomes
\begin{align*}
&\lim_{h\rightarrow\infty}\bra{q} 
{N}^{h}\big(\sum_{r\in\mathcal{C}^J_f}\prod_{i\in 
J}N^{(r_{i})}\big){N}^{h} \ket{q}=|q\Sigma|^{2t+1}\\ 
&\Leftrightarrow\lim_{h\rightarrow\infty}\bra{q} 
{N}^{h}\big(\prod_{-t\ldots t}\sum_{\sigma\in 
q\Sigma}N^{(\sigma)}\big){N}^{h} \ket{q}=|q\Sigma|^{2t+1}\\ 
&\Leftrightarrow\lim_{h\rightarrow\infty}\bra{q} 
{N}^{h}O^{2t+1}{N}^{h} \ket{q}=|q\Sigma|^{2t+1}\\ 
&\textrm{since }\sum_{\sigma\in 
q\Sigma}N^{(\sigma)}_{xy}=\sum_{\sigma\in 
q\Sigma}|\bra{\sigma}\delta\ket{xy}|^2=1\\ &\textrm{by the 
normalization condition.}
\end{align*}
Since $O^{2t+1}=|q\Sigma|^{2t}.O$ we have our condition. 
\hfill$\Box$

\subsection{Proof of Proposition \ref{finiterowcheck}} \label{pffiniterowcheck}

Since $\lim_{h\rightarrow\infty}\bra{q} 
{N}^{h}O{N}^{h}\ket{q}= \sum_{r\in\mathcal{C}^{[0,0]}_f} 
||\Delta^{\dagger}\ket{r}||^2$ this comes as a corollary of
Proposition \ref{localunitcheck}.\hfill $\Box$

\subsection{Proof of Proposition \ref{minbordervecs}.} \label{pfminbordervecs}

We have that 
$\ora{l}.\ora{r}=\ora{l}_q=\ora{l}_q=1$ since all three are 
by definition equal to $\lim_{h\rightarrow\infty}\bra{q} 
{N}^{h}\ket{q}=||\Delta^{\dagger}\ket{\ldots qqq 
\ldots}||^2=1$.(That the norm of the all-quiescent row is 
one stems from the fact that $\bra{q}\delta\ket{qq}=1$ and 
$\Delta$ is norm-preserving.) All entries in $\ora{l}$ and 
$\ora{r}$ are nonnegative because all entries in $N$ and 
$\ket{q}$ are nonnegative.\\ Now note that
\begin{align*}
(\sum_i \ora{l}_i)(\sum_i
\ora{r}_i)&=\lim_{h\rightarrow\infty}\bra{q}
{N}^{h}O{N}^{h}\ket{q}\\ &= 
\sum_{r\in\mathcal{C}^{[0,0]}_f} 
||\Delta^{\dagger}\ket{r}||^2\leq|q\Sigma|
\end{align*}
where the last line was obtained using $\sum_{\sigma\in 
q\Sigma}N^{(\sigma)}_{xy}=1$ and Lemma \ref{lemnpfs}. (It 
is clear that this inequality is saturated if and only if 
$\Delta$ has unit rows, since this was the content of 
proposition \ref{localunitcheck}.) Using this and 
nonnegativity we have $(\sum_i \ora{r}_i)=\ora{l}_q(\sum_i 
\ora{r}_i)\leq(\sum_i \ora{l}_i)(\sum_i \ora{r}_i)\leq 
|q\Sigma|$, hence the entries in $\ora{r}$ are finite, and 
symmetrically so for $\ora{l}$.\\ Say $\exists 
x\in\Sigma,\; [\ora{l}_x=0\;\wedge\;\ora{r}_x=0]$. Then 
$\bra{\ldots qq \ldots}\Delta\ket{\ldots qqxqq \ldots}\neq 
0$, which contradicts the fact that 
$\bra{q}\delta\ket{qq}=1$ and $\Delta$ is norm-preserving, 
and yields condition (iii). Say $\exists x,y\in\Sigma,\; 
[N_{xy}\neq 0\wedge\ora{l}_x=0\;\wedge\;\ora{l}_y=0]$. Then 
$\bra{\ldots qq \ldots}\Delta\ket{\ldots qqxyqq \ldots}\neq 
0$, which again contradicts the fact that 
$\bra{q}\delta\ket{qq}=1$ and $\Delta$ is norm-preserving, 
and yields condition (iv).\\ The fact that $\ora{r}$ is an 
eigenvector of $N$ is trivial, since 
$\ora{r}=\lim_{h\rightarrow\infty}{N}^{h}\ket{q}=\lim_{h\rightarrow\infty}{N}^{h+1}\ket{q}=N\ora{r}$. 
We now show it is minimal in $S=\{v\;|\;\; \mathbf{0}\leq 
v\;\wedge\; Nv=v\;\wedge\;v_q=1\}$. Say $w\in S$. We have 
$w=\ket{q}+u$, with $u$ a vector having nonnegative 
entries. Hence
\begin{align*}
w&=\lim_{h\rightarrow\infty}{N}^{h}w\\ &= 
\lim_{h\rightarrow\infty}{N}^{h}\ket{q}+\lim_{h\rightarrow\infty}{N}^{h}u\\ 
&=\ora{r}+\lim_{h\rightarrow\infty}{N}^{h}u.
\end{align*}
But since $N$ and $u$ have nonnegative entries, 
$\lim_{h\rightarrow\infty}{N}^{h}u$ has nonnegative entries 
and $\ora{r}\leq w$. Symmetrically so for 
$\ora{l}$.\hfill$\Box$

\end{document}